\documentclass[aps,onecolumn,superscriptaddress,floatfix,longbibliography,10pt,notitlepage,nofootinbib]{revtex4-1}

\usepackage{lipsum}
\usepackage{graphicx}
\usepackage[export]{adjustbox}
\usepackage{amsmath,amssymb,wasysym}
\usepackage{braket}
\usepackage{mathtools}
\usepackage{mathrsfs}
\usepackage{verbatim}
\usepackage{makecell}
\usepackage{cases}
\usepackage{bbm}
\usepackage{enumitem}
\usepackage[dvipsnames]{xcolor}
\usepackage{hyperref}

\newcommand{\im}{\mathbf{i}}

\newcommand{\Pstr}{\mathbf{s}}
\newcommand{\abs}[1]{\left| #1\right|}

\newcommand{\Tr}{\mathrm{Tr}}

\newcommand{\BigO}{\mathcal{O}}

\newcommand{\sign}{\mathrm{sign}}

\begin{document}
\title{A statistical mechanism for operator growth}
\author{Xiangyu Cao}
\address{Laboratoire de Physique de l'Ecole Normale Sup\'erieure, ENS, Universit\'e PSL,
CNRS, Sorbonne Universit\'e, Universit\'e de Paris, 75005 Paris, France}
\email{xiangyu.cao@ens.fr}
\begin{abstract} 
   It was recently conjectured that in generic quantum many-body systems, the spectral density of local operators has the slowest  high-frequency decay as permitted by locality. We show that the infinite-temperature version of this  ``universal operator growth hypothesis'' holds for the quantum Ising spin model in $d \ge 2$ dimensions, and for the chaotic Ising chain (with longitudinal and transverse fields) in one dimension. Moreover, the disordered chaotic Ising chain that exhibits many-body localization can have the same high-frequency spectral density asymptotics as thermalizing models. Our argument is statistical in nature, and is based on the observation that the moments of the spectral density can be written as a sign-problem-free sum over paths of Pauli string operators. 
\end{abstract}
\maketitle

\section{Introduction}
A basic question of many-body quantum dynamics at nonzero energy density is how to characterize generic thermalizing systems and distinguish them those which fail to thermalize completely. The very influential approach of ``quantum chaos'', initiated around 1980~\cite{casati,berry-tabor,BERRY1981163,bohigas}, focuses on the correlation between energy levels of the Hamiltonian that generates the dynamics. Generic systems are expected to behave like random matrices, and exhibit level repulsion. A related approach, known as the eigenstate thermalization hypothesis~\cite{deutsch91,srednicki94,rigol2008thermalization,deutsch2018eigenstate}, concerns the matrix element of local operators in the energy basis.  

Recently, there has been a renewed interest in this question, with an emphasis on the growth of local operators under Heisenberg time evolution~\cite{nahum2018operator,khemani2018operator,von2018operator,chan18chaos}. Indeed, if $O$ is a local observable in a many-body system, $O(t) = e^{\im t H} O e^{-\im t H} $ will evolve into a complex many-body object under a generic interacting Hamiltonian $H$ as $t$ becomes large; in a non-interacting system, there is no such a growth of complexity. The nature of the dynamics manifests itself in the extent of operator growth. Yet, the latter cannot be easily captured by correlators such as $\Tr[O(t) O]$, because they can have similar decay behaviors in both generic and free systems. Unusual quantities~\footnote{Technically speaking, the quantities are unusual in that they are defined on a multi-folded Keldysh contour.}, such as ``out of time order'' correlators~\cite{larkin,Maldacena:2015waa}, or operator entanglement~\cite{dubail,Nie_2019}, have been proposed as quantitative probes of operator growth and  quantum thermalization. 

Nevertheless, it was observed~\cite{hypothesis} that the high frequency regime of the spectral density, which is the Fourier transform of the infinite-temperature auto-correlation function
\begin{equation}
    \Phi(\omega) = \int_{-\infty}^{\infty} \frac{1}{\Tr[1]} \Tr[O(t) O]   e^{\im \omega t} d t  
\end{equation}
can capture operator growth. The observation is based on a sequence of mathematical relations, from the high-frequency tail of $\Phi(\omega)$, to the asymptotics of the Lanczos coefficients~\cite{Lubinsky1987,avdoshkin}, and the growth of the ``K(rylov)-complexity''~\cite{hypothesis,rabinovici,Rabinovici:2020ryf}. The latter provides in turn an upper bound on out of time order correlators (OTOCs)~\cite{hypothesis,murthy19,avdoshkin}. In this context, the authors of \cite{hypothesis} put forward a ``universal operator growth hypothesis'' according to which the spectral density decays as an exponential 
\begin{equation}
\label{eq:expoential}   \Phi(\omega) \sim e^{-|\omega|/\omega_0} \text{   (generic)}
\end{equation} 
in generic thermalizing systems with a log correction in one dimension. Note that the term ``universal'' refers to the exponential form, but not on the exponent $\omega_0$, which depends on the Hamiltonian $H$ (and also $O$, to a lesser extent). It is shown in \cite{hypothesis} that the exponential tail corresponds to an exponential growth of the K-complexity $\sim e^{ \pi \omega_0  t}$. In large $N$ and semi classical systems, that provides a tight bound on the Lyapunov exponent characterizing the exponential growth of OTOCs. Moreover, unlike OTOCs, the K-complexity can have extended exponential growth away from these limits, providing a general quantitative measure of operator complexity growth.

We mention in passing that, besides the relation to operator growth, the high-frequency tail of the spectral density is also directly related to the slow heating (pre-thermalization) of a system under a fast periodic drive $\delta H \propto O \cos(\omega t)$~\cite{abanin15,abanin17}, which can be measured experimentally~\cite{bloch2020heating}. The spectral density also governs the off-diagonal matrix elements in the eigenstate thermalization hypothesis.

The universal operator growth hypothesis has been justified heuristically. The main pieces of evidence are analytic calculations in the large $N$ Sachdev-Ye-Kitaev~\cite{SY93,kitaev15} model, numerics on 1d spin chains, and a general upper bound~\cite{abanin15} 
\begin{equation}
     \Phi(\omega) \lesssim e^{-|\omega|/\omega_{\max}}   \,,
\end{equation}
for some $\omega_{\max}$ which is given by the local bandwidth of the Hamiltonian $H$ times some geometric factor of order $1$~\cite{abanin15,hypothesis}; for an explicit example, see \eqref{eq:upper_bound_1d} below. In other words, the exponential decay is the slowest possible. However, showing that that generic systems satisfy a similar \textit{lower} bound,
\begin{equation}
    \Phi(\omega) \gtrsim e^{-|\omega|/\omega_{\min}} \,, \label{eq:lower_bound_intro}
\end{equation}
with possibly a different exponent $\omega_{\min} > 0$, is harder (the problem of precisely pinpointing the exponent $\omega_0$ for a given finite-dimensional model is even harder, and beyond the scope of this work). To our knowledge, away from large $N$ and classical limits, it has been only done for a rather artificial spin model in 2d~\cite{bouch} (see \eqref{eq:H_bouch} below), but not for any standard chaotic systems. For example, the 1d Ising model with transverse and longitudinal fields (also known as the ``chaotic Ising chain''), 
\begin{equation}
    H = \sum_{j} [J Z_j Z_{j+1} + h_X X_j + h_Z Z_j ]  \label{eq:H_Ising}
\end{equation}
is believed to satisfy the hypothesis (the 1d version, see Section~\ref{sec: liouville} below), but so far this is only a numerical observation~\cite{hypothesis}. 

In this work, we present an argument that allows to show such a lower bound as \eqref{eq:lower_bound_intro} for a class of spin models, including \eqref{eq:H_Ising} and its higher-dimensional analog.  The main idea is the following. The high frequency asymptotics of the spectral density is given by that of its moments 
\begin{equation}
  \mu_{2n} = \int \Phi(\omega)\omega^{2n} \frac{d \omega }{2\pi } \,.
\end{equation}
In particular, the lower bound \eqref{eq:lower_bound_intro} of interest amounts to the following lower bound on the moments:
\begin{equation}
    \mu_{2n} \gtrsim \left( \frac{2n \omega_0 }{e} \right)^{2n} \;  \Leftrightarrow  \; \Phi(\omega) \gtrsim e^{-\omega/|\omega_0|} \,. \label{eq:moment_Phi}
\end{equation}
Showing such a lower bound on the moments will be the main focus of this work.
For this, we represent them as  ``partition sums'' over paths in a graph of Pauli strings connected by the interaction term in the Hamiltonian. The general upper bound on $\mu_{2n}$, and thus on $\Phi(\omega)$, is obtained by estimating the total number of paths, which is constrained by locality (or $k$-locality in all-to-all models). For the lower bound, one needs to take into account cancellation between paths with positive and negative contributions. However, this issue is eliminated for a large class of models that we shall identify, including~\eqref{eq:H_Ising}: the sum is sign-problem-free. This allows us establish lower bounds on the moments by counting. As a result, we verify the universal operator growth hypothesis in the chaotic Ising chain as well as its analog in 2d (in which case the longitudinal field is not necessary). 

Furthermore, this method also applies to disordered spin chains, and allows us to address a pending issue regarding the operator growth in many-body localized (MBL) systems~\cite{mbl_rev}. Since MBL prevents thermalization, one would expect that that the universal operator growth hypothesis does not apply, and $\Phi(\omega)$ decays qualitative faster, say $\sim e^{-\omega^2 / \omega_0^2}$~\cite{khait16mbl}, as suspected in integrable models~\cite{hypothesis,rigols19}. We will show that this is \textit{not} the case for MBL systems; instead, they behave in the same way as generic thermalizing systems as far as $\Phi(\omega \to \infty)$  is concerned. 

\section{Moments and Liouvillian graph}\label{sec: liouville}
Throughout this paper, we consider models of spin-half's in the thermodynamic limit. Denoting the Pauli operators acting on the spin $i$ as $X_i, Y_i$ and $Z_i$, a basis of the operator Hilbert space is given by the ``Pauli strings'', which are finite products of single-site Pauli's, such as $X_i$, $Z_i Y_j$, etc. This basis is also orthonormal under the  infinite-temperature inner product 
\begin{equation}
    ( A \vert B ) = \frac{\Tr[A^\dagger B]}{\Tr[1]} \,.
\end{equation}

We are interested in the growth of a simple Pauli string operator $O$  under the Heisenberg dynamics generated by a local Hamiltonian $H$, which is an infinite sum of Pauli strings of bounded length, with bounded coefficients. An example is the chaotic Ising chain \eqref{eq:H_Ising}. Given $H$ and $O$, the moments $\mu_{2n}$ can be computed as follows:
\begin{equation}
    \mu_{2n} =  ( O \vert  \mathcal{L}^{2n} O )  \,,\,  \label{eq:mu2nL}
\end{equation}
where $\mathcal{L}$ is the Liouvillian:
\begin{equation}
    \mathcal{L} A := [H, A]  \,.
\end{equation}
Note that $\mu_{2n}$ is the norm squared of $ \mathcal{L}^{n} O $, so is always positive. 
Now, expanding the matrix product $ \mathcal{L}^{2n} $ in the Pauli string basis, we obtain
\begin{equation}
    \mu_{2n} = \sum_{O = \Pstr_0, \Pstr_1, \dots, \Pstr_{2n-1}, \Pstr_{2n} = O} 
    \prod_{k= 0}^{2n-1} ( \Pstr_{k+1} \vert  \mathcal{L}^{2n} \vert  \Pstr_{k} ) \,, \label{eq:pathsum}
\end{equation}
where $\Pstr_k$'s are Pauli strings. Namely, the $2n$-th moment is a sum over paths of length $2n$ in a ``Liouvillian graph''. Its vertices are labelled by Pauli strings, and are connected by the non-zero matrix elements of the Liouvillian. The contribution of a path is given by the product of matrix elements of $\mathcal{L}$ along the edges, just like in a path integral. The endpoints of the paths are fixed by the operator $O$. For example, if $H$ is given by \eqref{eq:H_Ising} and $O = Z_0$. Then the following path
\begin{equation}
    (\Pstr_0, \dots, \Pstr_6) = (Z_0, Y_0, X_0 Z_1, Y_0 Z_1, X_0, Y_0, Z_0) 
\end{equation}
contributes an amplitude $2^{6} (h^X)^2 J^2 (h^Z)^2$ to the moment $\mu_{6}$ (recall that $[X, Y] = 2\im Z$ and so on). 

Using the path-sum formula~\eqref{eq:pathsum}, it is not hard to derive an upper bound for the moments, which was done previously~\cite{araki1969gibbs,abanin15,hypothesis}. Indeed, the number of paths in \eqref{eq:pathsum} is constrained by locality: at the $k$-th step, the Pauli string has a support of size $\BigO(k)$, so it has non-zero commutator with only $\BigO(k)$ terms in the Hamiltonian, and thus we have only $\BigO(k)$ choices of $\Pstr_{k+1}$ given $ \Pstr_{k+1}$.
The amplitudes of the paths are bounded uniformly by some $J_{\max}^{2n}$ which corresponds to the maximal coupling constant in the $H$. Then, one can bound \eqref{eq:pathsum} term by term, and obtain
\begin{equation}
    \mu_{2n} \le J_{\max}^{2n} (2n)! \sim n^{2n} e^{C n} 
\end{equation}
for some constant $C$ independent of $n$. For 1d short-range Hamiltonians, the bound can be improved by a log correction~\cite{araki1969gibbs,bouch}:
\begin{equation}
      \mu_{2n} \le \frac{{n}^{2n}}{(\ln n)^{2n}} e^{\BigO(n)}  \,, \label{eq:mu2n-log}
\end{equation}
which corresponds to $\Phi(\omega) \lesssim e^{- \frac{ |\omega|}{\omega_0} \ln |\omega| }$ in terms of spectral density decay. The one dimension is special essentially because the Pauli string can only grow at the boundary, which significantly limits the number of paths. 

\section{Positive Liouvillian graphs}\label{sec:positive}
Lower bounds on $\mu_{2n}$ are less trivial to obtain. \textit{A priori}, we would need to account for non-positive amplitudes and their destructive interference. Like the sign problem in quantum Monte Carlo, this can be hard in general, yet one can make progress by avoiding it. Let us define that a Hamiltonian is \textit{positive} if there exists a choice of gauge, i.e., a phase $\theta_{\Pstr}$ for each Pauli string, that makes all the matrix elements of the Liouvillian non-negative 
\begin{equation}
     ( \tilde\Pstr' \vert  \mathcal{L}^{2n} \vert  \tilde\Pstr )  \ge 0 \,,\, \text{where } \tilde\Pstr := \Pstr e^{\im \theta_{\Pstr}} \,
\end{equation}
are Pauli string operators with the phase attached. The choice of gauge is unique in every connected component of the Liouvillian graph; hence having a positive Liouvillian graph or not is essentially a property of the Hamiltonian.  It is clear that, if $H$ is positive, the path sum of the moments~\eqref{eq:pathsum} will be sign-problem-free: all the terms are non-negative. 

The main observation of this work is that, the following Hamiltonian
\begin{equation}
    H = \sum_{ij}  J_{ij} Z_i Z_j  + \sum_i [ h^Z_i Z_i + h^X_i X_i]  \label{eq:H_gen}
\end{equation}
is positive, provided for any $i$, there exists $s_i \in \{1,-1\}$ such that for any $j \ne i$
\begin{equation}
    s_i J_{ij} h_j^Z \ge 0 \,. \label{eq:condition}
\end{equation}
Note that any homogeneous model, i.e., one with $J_{ij} = J$, $h_i^X = h_X$, $ h_i^Z = h^Z$, satisfies the above condition and is thus positive. In particular, the chaotic Ising chain~\eqref{eq:H_Ising} is positive.

 To prove the above result, we claim that the matrix elements of the Liouvillian are non-negative between the following Pauli strings with a phase:
\begin{align}
  &  \tilde\Pstr = \prod_{i} S_i \,,\, S_i \in \{ I,  \tilde{X}_i,  \tilde{Y}_i, \tilde{Z}_i \} \text{ where }\\ 
   & \tilde{X}_i := - s_i \sign(h^X_i h^Z_i)  X_i \,,\, \tilde{Y}_i := -\im s_i \sign(h^X_i) Y_i \,,\, \tilde{Z}_i  = s_i Z_i \,. \label{eq:gauge_choice}
\end{align}
That is, the gauge is chosen by attaching a phase to each Pauli matrix in the string. To see why the above gauge choice works, let us observe that the nonzero matrix elements can be divided into three classes, corresponding to the three types of terms in $H$: 
\begin{enumerate}
    \item $\Pstr$ and $\Pstr'$ differ only in one site, $i$, where $S_i = \tilde{X}_i$ and $S'_i = \tilde{Y}_i$ (or vice versa). Then, only the term $h_i^Z Z_i$ is contributing to the matrix element, and we have
    \begin{eqnarray*}
        & ( \tilde\Pstr' \vert  \mathcal{L} \vert  \tilde\Pstr ) 
         =
         \left(\tilde{Y}_i \vert [h_i^Z Z_i, \tilde{X}_i] \right) \\
         =& \im s_i  \sign(h_i^X h_i^Z) (-s_i \sign(h_i^X) ) h_i^Z  \left({Y}_i \vert [ Z_i, {X}_i] \right)  \\
         =& 2|h_i^Z|  \ge 0 
    \end{eqnarray*} 
    by \eqref{eq:gauge_choice} and $[Z, X] = 2\im Y$. 
    \item A similar analysis applies to the case where $h^X_i X_i$ is contributing, and $\Pstr$, $\Pstr'$ differ in that $S_i = \tilde{Y}_i$ and $S'_i = \tilde{Z}_i$ or vice versa. 
    \item $\Pstr$ and $\Pstr'$ differ only in two sites, $i$ and $j$, and $J_{ij} Z_i Z_j$ is the only term that contributes to the matrix element:
    $ [J_{ij } Z_i Z_j, \Pstr] \propto \Pstr' \,. $ For $\Pstr$ to not commute with $Z_i Z_j$, we must have $$ \left( S_i \in \{I, \tilde{Z}_i\} \,,\, S_j \in \{\tilde{X}_j, \tilde{Y}_j\} \right) \text{ or } \left( S_i \in\{\tilde{X}_i, \tilde{Y}_i\}  \,,\, S_j \in \{I, \tilde{Z}_j\} \right) \,.$$
  Then it is straightforward to verify that
    $$  ( \tilde\Pstr' \vert  \mathcal{L} \vert  \tilde\Pstr) = \begin{cases}  2 J_{ij} \sign(h_j^Z) s_i & S_i \in \{I, Z_i\}  \\ 
     2 J_{ij} \sign(h_i^Z) s_j & S_j \in \{I, Z_j\} \end{cases} $$
     which is non-negative in any case, thanks to the condition \eqref{eq:condition}.
\end{enumerate}
Thus, we have shown that all the matrix elements of $\mathcal{L}$ are positive with the gauge choice~\eqref{eq:gauge_choice}, and concluded the demonstration. 

Before proceeding, we remark that interacting Hamiltonians are rarely positive. For example, one can check that the XXZ chain $H = \sum_{j} [X_j X_{j+1} + Y_j Y_{j+1} + \Delta Z_j Z_{j+1}]$ is not positive for $\Delta \ne 0$, although it is positive with $\Delta = 0$. The chaotic Ising model with an additional $X_j X_{j+1}$ interaction is not positive either. The existence of the class \eqref{eq:H_gen}, \eqref{eq:condition} is not trivial.

\section{Lower bound for moments}\label{sec:proof}
We now apply the positiveness result of the previous section to show lower bounds on the moments in a few models in the family defined by \eqref{eq:H_gen} and \eqref{eq:condition}. The general strategy is to use the path-sum formula~\eqref{eq:pathsum} for the moments; by positiveness, all paths have a nonnegative contribution, and a sum over any subset of paths $\mathcal{P}$ gives a lower bound: $$ \mu_{2n} \ge \sum_{\mathcal{P}}   \prod_{k= 0}^{2n-1}\abs{ ( \Pstr_{k+1} \vert  \mathcal{L}^{2n} \vert  \Pstr_{k} ) } \,. $$
(We can take the absolute value because the product is guaranteed to be positive.) 
It remains to choose a suitable subset $\mathcal{P}$ over which the path sum is convenient to estimate. 

\subsection{Chaotic Ising chain}
Let us first consider the 1d chaotic Ising chain~\eqref{eq:H_Ising}, which is positive. For concreteness, let the operator be $O = Z_1$ (although the reasoning below applies more generally).
The subset of paths will be those which consist of three parts:
\begin{enumerate}
    \item An initial growth
\begin{align}
    Z& \stackrel{X}\to Y \stackrel{ZZ}\to XZ \stackrel{X}\to XY \stackrel{ZZ}\to XXZ \to XXY \to XXXZ \to \dots \to \underbrace{X\dots X}_{k}Z \,, \label{eq:growth}
\end{align}
where the notation should explain itself: $Z = Z_1$, $XXY = X_1 X_2 Y_3$, etc. This part of the path corresponds to $2k$ applications of the Liouvillian. 
\item A scrambling regime
\begin{equation}
    \underbrace{X\dots X}_{k}Z \to \dots \to \underbrace{X\dots X}_{k}Z\,, \label{eq:1dscrambling}
\end{equation}
where all the matrix elements correspond to the commutation with some $Z_i, i = 0, \dots, k - 1$ on one of the sites of the $ \underbrace{X\dots X}_{k}$ string. It is not hard to show that, the number of such paths~\eqref{eq:1dscrambling} with length $2\ell$, denoted $N_{k,\ell} $ is given by the following:
\begin{equation}
    N_{k,\ell} = \sum_{j=0}^k \binom{k}{j} (k-2j)^{2 \ell } 2^{- k} \ge k^{2 \ell } 2^{-k} \,. 
\end{equation}
\item A final return, where we undo the growth process of step (i) \eqref{eq:growth}. 
\end{enumerate}
In summary, there are $N_{k,\ell}$ paths, each of which contributes to the moment $\mu_{2n}$, with $n = \ell + 2k$, the following amplitude 
\begin{equation}
     \prod_{k} \abs{ ( \Pstr_{k+1} \vert  \mathcal{L}^{2n} \vert  \Pstr_{k} ) }  = 
   2^{2n} (J h_X)^{2k} h_Z^{2\ell} \,.
\end{equation} 
Therefore, we have 
\begin{equation}
   \mu_{2n} \ge 2^{2n} (J h_X)^{2k} h_Z^{2\ell} k^{2 \ell } 2^{- k}   \,,\, n = \ell + 2k \,.
\end{equation}
To get a best bound, we can optimize $N_{k, \ell= n-2k}$ over $k$. For large enough $n$, it is not hard to show that the optimum is given by $k \approx n / (2\ln n)$, and that
\begin{equation}
    \mu_{2n} \ge \left( \frac{2 n \omega_{\min}}{e \ln n} \right)^{2n} e^{\BigO(n / \ln n)}
\,,\, \omega_{\min} = |h_Z| / 2 \,. \label{eq:lower_bound_1d}
      \end{equation}
The asymptotics of the right hand side is the same as \eqref{eq:moment_Phi}, except for the log correction $(\ln n)^{-2n}$. This log correction is necessarily present, because a similar asymptotics as \eqref{eq:lower_bound_1d} (with a different $\omega_{\max}$ in lieu of $\omega_{\min}$) is also an upper bound for general 1d short-range spin models. Indeed, by the argument of \cite{hypothesis}, Appendix F, one can show that 
\begin{equation}
    \mu_{2n} \le \left( \frac{2 n \omega_{\max}}{e \ln n} \right)^{2n} e^{\BigO(n / \ln n)} \,,\, \omega_{\max} = 2 \max(|J|, h_X, h_Z) \,.\label{eq:upper_bound_1d}
\end{equation}
The moments are bounded from both sides by the same asymptotic form, with different exponents $\omega_{\min}$ and $\omega_{\max}$. Therefore, we can conclude that the chaotic Ising chain is compatible with the universal operator growth hypothesis in 1d, with $\omega_0 \in [\omega_{\min}, \omega_{\max}]$. Of course, the lower bound is nontrivial only if $h_Z\ne 0$, as well as $h_X$ and $J$. When $h_Z = 0$, the model becomes equivalent to a free Majorana fermion model by a Jordan-Wigner transform, and the moments are known to grow much slower $\mu_{2n} \sim n! \ll \left({n}/{\ln n}\right)^{2n}$~\cite{vishwanath1990recursion}, see also Appendix~\ref{app}. 

\subsection{Many-body localization}
Spin chains with a strong disordered magnetic field are believed to exhibit many-body localization (MBL). This is a distinct dynamical phase at nonzero energy density, where thermalization is hampered by the existence of an extensive number of local integrals of motion~\cite{ROS2015420}. While much of the evidence and theory of MBL are numerical and phenomenological, Imbrie~\cite{imbrie} managed to rigorously establish the existence of MBL for strongly disordered chaotic Ising models, under a mild level statistics assumption. The theorem applies in particular to random Hamiltonians 
\begin{equation}
     H = \sum_{j} [J_j Z_j Z_{j+1} + h^X_j X_j + h^Z_j Z_j ] 
\end{equation}
where $J_j \in [J, 2J]$, $h^X_j \in [h_X, 2 h_X]$ and $ h^Z_j \in [h_Z, 2 h_Z] $ are independent and uniformly distributed, provided $h_Z \gg J, h_Z \gg h_X$ (strong disorder). Now, if $J, h_X, h_Z$ are all positive, $H$ will be always positive since the condition \eqref{eq:condition} is satisfied. Then, the argument of the previous section immediately implies that the moments satisfy the same lower bound \eqref{eq:lower_bound_1d} (and a similar upper bound). 

This result should not be interpreted as questioning the existence of MBL in this model (see however \cite{vsuntajs2019quantum,abanin2019distinguishing,Panda2020mbl,piotre} for a recent debate on this issue). Rather, it shows that the high-frequency behavior of the spectral function is rather blind to MBL. This is probably because $\Phi(\omega)$ sums over all the matrix elements between energy eigenstates with energy difference $\omega$, while to detect MBL necessitates looking at fluctuations of matrix element amplitudes~\cite{serbyn}.

\subsection{Higher dimensions}
Generic spin models in two and higher dimensions are believed to satisfy the universal operator growth hypothesis without the log correction, i.e., the moments should have the following asymptotic lower bound
\begin{equation}
    \mu_{2n} \ge n^{2n} \exp(-\BigO(n)) \,. \label{eq:mu2n_lower_2d}
\end{equation}
This has been explicitly proven only in a spin model on a square lattice considered by Bouch~\cite{bouch}:
\begin{equation}\label{eq:H_bouch}
H =
\sum_{x,y}
\left( X_{x,y} Z_{x+1,y} + Z_{x,y} X_{x,y+1} \right) \,,
\end{equation}
which is rather exotic from a physical point of view. 

Here, we show that \eqref{eq:mu2n_lower_2d} is satisfied by the quantum Ising model in 2d, which is experimentally realizable~\cite{TFIM1,TFIM2,tfim3}. (The same statement in higher dimensions follows immediately.) We consider the Hamiltonian 
\begin{equation}
   H = \sum_{\left<ij\right>} J Z_i Z_j + \sum_i h_X X_i \,,   \label{eq:Ising2d}
\end{equation}
where $\left< ij \right>$ denotes nearest-neighbor sites on a square lattice. For convenience, we let $J = h_X = 1$ and take the initial operator to be $Z_{0,0}$, but the argument can be easily adapted to other cases. 

By the result of Section~\ref{sec:positive}, the Hamiltonian~\eqref{eq:Ising2d} is positive. Since any path of length $2n$ contributes to $\mu_{2n}$ an amplitude $2^{2n} \sim e^{\BigO(n)}$, bounding the moments from below reduces to counting the number of (a subset of) paths in the Liouvillian graph. Inspired by \cite{bouch}, we shall consider a set of paths of length $2n$, whose first $n$ steps build a lattice tree of size $m \sim n/2$ and the last $n$ steps demolish it. An example of the tree-building process is the following:
\begin{equation}
\begin{split}
&
    \begin{matrix}    
                        I & I & I \\
                        I & Z & I \\
                        I & I & I 
                     \end{matrix} \, \to
                       \begin{matrix}  
                      I & I & I \\
                        I & {\bf Y} & I \\
                        I & I & I  
                     \end{matrix} \,  \to 
                     \begin{matrix}  
                      I & I & I \\
                        I & \bf X & \bf Z \\
                        I & I & I  
                     \end{matrix} \to 
                       \begin{matrix}  
                      I & I & I \\
                        I & X & \bf{Y} \\
                        I & I & I  
                     \end{matrix} \to 
                     \begin{matrix}  
                      I & I & \bf Z \\
                        I & X & \bf X\\
                        I & I & I  
                     \end{matrix} \\ \to& 
                      \begin{matrix}  
                      I & I & Z \\
                        I & Y & X\\
                        I & \bf Z & I  
                     \end{matrix} 
    \to 
     \begin{matrix}  
                      I & I & Z \\
                        \bf Z & X & X\\
                        I & Z & I  
                     \end{matrix} \to 
                       \begin{matrix}  
                      I & I & Z \\
                        \bf Y & X & X\\
                        I & Z & I  
                     \end{matrix} \to 
                      \begin{matrix}  
                     \bf Z & I & Z \\
                       \bf  X & X & X\\
                        I & Z & I  
                     \end{matrix} \to \dots
                     \end{split} \label{eq:growth2d}
\end{equation}
where the bold faced Pauli's highlight the growth process. Reversing the arrows gives an example of a tree-demolishing process. Compared to step (i) in the 1d case~\eqref{eq:growth}, the growth in 2d~\eqref{eq:growth2d} is similar in that two steps are needed to build one new vertex: one from $I \to Z$ using $ZZ$,  and the other $Z \to Y$ using $X$, so that the vertex is ``activated'' and available to generate more $Z$'s. Now, the crucial difference from 1d is that there are lattice trees in 2d that can be grown by far more ways. More precisely, we claim that there are trees of arbitrarily large size $m$ that can be built by a number of ways that is
\begin{equation} 
N_m \gtrsim m^{2m} e^{-\BigO(m)}  \,. \label{eq:waystogrow}
\end{equation}
This would imply the desired lower bound on moments: since the building and demolition processes are independent, we would have
$$ \mu_{2n} \ge g_{m = n/2}^2 \sim n^{2n} e^{-\BigO(n)} \,, $$
for a sequence of arbitrarily large $n$.~\footnote{This is enough to verify the hypothesis, as it implies that the auto-correlation function has a pole in the imaginary axis~\cite{hypothesis,bouch}.}

To show the claim~\eqref{eq:waystogrow}, we use a \textit{tour-de-force} result of \cite{bouch} (Theorem 6.1, Lemme 6.2). It states that that there exists a constant $C > 0$, and rooted lattice trees with arbitrarily large number of vertices $m$, such that 
\begin{equation}
    \frac{m!}{\prod_{e} w(e)} \ge \frac{ m!}{ C^m} \,
\end{equation}
where $w(e)$ is the number of descendants of an edge $e$, and the product is over all the edges of the tree. The LHS is equal to the ways of growing such a tree \textit{if the activation steps are not needed}. This latter fact is shown in \textit{ibid}, Lemma 6.2. The idea is that, the order of adding vertices is arbitrary (hence $m!$), modulo the constraint that for every sub-tree ($\prod_e$), its root must be built first, before the descendants (this gives a factor $1/w(e)$).
Now, with the activation steps, the above reasoning can be adapted to show that the number of ways of build the same tree in $2m$ steps is at least
\begin{equation}
    \frac{(2m)!}{\prod_{e} 2w(e) (2w(e) -1)} \ge \frac{ (2m)!}{ C^{2m}} \sim m^{2m} e^{-\BigO(m)}\,.
\end{equation}
So we have shown our claim above, and concluded the demonstration. 

We remark that the argument above does not provide a reasonable estimate of $\omega_0$ that characterizes the spectral density decay (the lower bound given by \cite{bouch} is too small). Nevertheless, given the positiveness of the Hamiltonian, it is in principle possible to estimate $\omega_0$ by a Monte Carlo sampling of paths. We will leave this to a future study.

\section{Conclusion}\label{sec:discussion}
We showed that a standard example of generic spin Hamiltonian --- the (chaotic) quantum Ising model --- satisfies with the universal operator growth hypothesis, in the sense that the moments satisfy a lower bound. Equivalently, the spectral density has the slowest possible high-frequency decay. Our demonstration is essentially rigorous, and makes a first step in putting the hypothesis on a more solid footing. 

The crux of our argument is that the moments in these models are given by a ``sign-problem-free'' sum of paths: somehow, the coherent interference effects in quantum dynamics disappear when it comes to the high-frequency regime of correlation functions: instead, it acquires a ``statistical'' nature. Note that in the large $N$ Sachdev-Ye-Kitaev model, the moments also have a combinatorial interpretation, in terms of counting ``melon'' Feynman diagrams. It is rather unexpected that the same happens in a family of small $N$, finite dimensional spin models, where the diagrams are intractable. While the family of positive Hamiltonians we considered are ``fine-tuned'', positiveness is just one way to express the moments as a sign-problem-free sum. It will be interesting to see whether there are other ways to do so for more general Hamiltonians.

Another surprising outcome of our argument is that the high-frequency regime of the spectral density does not distinguish many-body localization from complete thermalization. This means that the K-complexity is a different from level statistics as probe of ``quantum chaos'', even for systems away from the classical limit (in the latter limit, this is known to be the case~\cite{xu2020}). So far, the only nontrivial interacting systems that may have a qualitatively slower K-complexity growth are Bethe-Ansatz integrable models, however this claim is only supported by numerics~\cite{hypothesis,rigols19}. Analytically identifying interacting models with non-generic K-complexity growth is an outstanding question for future study.

\textbf{Acknowledgements.} I thank Ehud Altman, Yimu Bao, Daniel Parker, and especially Alexandre Avdoshkin, for stimulating discussions and collaboration on related projects. I acknowledge support from a United States Department of Energy grant DE-SC0019380 during my postdoctoral appointment at University of California, Berkeley,

\appendix
\section{Moments in quantum Ising model}\label{app}
In this appendix, we recall the calculation of the moments in the (integrable) quantum Ising chain, of an operator that is non-local in the fermionic picture. 

For simplicity let's consider the quantum Ising model at criticality:
\begin{equation}
    H = \frac12 \sum_j [Z_j Z_j + X_j] \,.
\end{equation}
By Jordan-Wigner transform it is equivalent to a Majorana chain 
\begin{equation}
    H = \sum_{j \in \mathbf{Z}/2} \im \gamma_j \gamma_{j+1/2}  \,,\, \{ \gamma_i , \gamma_j\} = \delta_{ij} \,.
\end{equation}
Every Pauli string in the spin model is mapped to a Majorana string. For instance, The operator $Z_0$ (of which we consider the moments) becomes a semi-infinite Majorana string:
\begin{equation}
    Z_0 = \dots  \gamma_{-3}  \gamma_{-2} \gamma_{-1} \gamma_0  = 
    \vert \dots \bullet \bullet \bullet \bullet \circ \circ \circ \circ  \dots )
\end{equation}
where a $\bullet$ is a site with Majorana and $\circ$ without. 
 
The action of the Liouvillian on it generates nearest-neighbor hopping of the Majorana fermions subject to the simple exclusive constraint. For example 
\begin{equation}
    \mathcal{L} \vert \dots \bullet \bullet \bullet \bullet \circ \circ \circ \circ  \dots ) = \vert \dots \bullet \bullet \bullet  \circ \bullet \circ \circ \circ  \dots ) 
\end{equation}
while 
\begin{equation}
    \mathcal{L} \vert \dots \bullet \bullet \bullet  \circ \bullet \circ \circ \circ  \dots ) = \vert \dots \bullet \bullet \bullet  \circ  \circ \bullet \circ \circ  \dots ) +  \vert \dots \bullet \bullet  \circ \bullet \bullet   \circ  \circ \circ  \dots ) + 
    \vert \dots \bullet \bullet  \bullet \bullet  \circ  \circ  \circ \circ  \dots ) \,.
\end{equation}
Thanks to positiveness, we do not need to keep track of the phases, i.e., we can make appropriate choice of phases so that all the hopping amplitudes are $1$. 

The algebra of such hopping generators are well-known. Indeed, let 
\begin{equation}
    \mathcal{L} = \mathcal{A} + \mathcal{A}^\dagger
\end{equation}
where $ \mathcal{A}^\dagger$ is generates only hopping to the right and $ \mathcal{A}$ to the left. It is not hard to check that two terms form a harmonic-oscillator algebra:
\begin{equation}
    [ \mathcal{A},  \mathcal{A}^\dagger ] = 1 \,.
\end{equation}
restricted on the space accessible to $X$ (the space of partitions). Also note that $\vert X )$ corresponds to the vacuum in the oscillator analogy. So we have by Wick theorem
\begin{equation} 
\mu_{2n} =  (X \vert  \mathcal{L}^{2n}  \vert X ) = \frac{(2n)!}{n! 2^n} \sim n^n \exp(\BigO(n)) \,.
\end{equation}
This is qualitatively slower than $(n / \ln)^{2n} e^{\BigO(n)}$ in generic 1d chains, but faster then an exponential growth $  e^{\BigO(n)}$ which occurs for operators local in the fermionic picture such as $X_0$. 

\bibliography{refs}

\end{document}